\documentclass[preprint2]{aastex}
\usepackage{natbib}
\usepackage{layout}
\usepackage[latin1]{inputenc}
\usepackage{amssymb}
\usepackage{lscape}
\usepackage{hyperref}

\shorttitle{A new photopolymer based VPHG for astronomy}
\shortauthors{Zanutta et al.}

\begin{document}

\title{A new photopolymer based VPHG for astronomy: The case of SN 2013fj}

\author{\scshape {Alessio Zanutta, Marco Landoni, Andrea Bianco}}
\affil{INAF - Osservatorio Astronomico di Brera, Via Emilio Bianchi 46, I-23807 Merate, Italy}

\author{\scshape {Lina Tomasella, Stefano Benetti, Enrico Giro}}
\affil{INAF - Osservatorio Astronomico di Padova, Vicolo dell'Osservatorio 5, I-35122 Padova, Italy}

\begin{abstract}
The spectroscopic studies of near infrared emission arising from supernovae allow to derive crucial quantities that could better characterise physical conditions of the expanding gas, such as the CaII IR HVF spectral feature. For this reason is mandatory to have Diffractive Optical Elements (DOEs) with a spectral coverage in the range 8000 - 10000 $\textrm{\AA}$ (for low z sources) combined with a reasonable Signal to Noise Ratio (S/N) and medium-low resolution. In order to cope with all of those requirements we developed a Volume Phase Holographic Grating (VPHG) based on an innovative photosensitive material, developed by Bayer MaterialScience.
We demonstrated the capabilities of this new DOE through observation of SN 2013fj as case study at Asiago Copernico Telescope where AFOSC spectrograph is available. 
\end{abstract}

\keywords{Supernovae Ia; extragalactic astronomy; volume phase holographic grating; photopolymers; grism}

\section{Introduction}
Astronomical spectrographs are key instrumentation to tackle the open issues in astronomy. One of the most important element along with the detector is the dispersing element. In the last 15 years, the Volume Phase Holographic Grating (VPHG) technology has gained a lot of interest in astronomical field and it has been used in some spectrographs \citep{baldry, andrea10, barden, pazder}. The reasons for such interest stem from the fact that these gratings show unique features, such as i) the high peak efficiency (up to 100\% theoretically) both at low and large dispersion; ii) the ease of performance tuning and customisation (each VPHG is a master grating). A VPHG consists in a thin layer of holographic material, usually dichromated gelatine \citep{andrea10, barden}, which is sandwiched between two glass windows. The phase of incident light is modified passing through the gratings thanks to a periodic modulation (usually sinusoidal) of the refractive index written in the holographic material.
Hence the fundamental parameters that rule the overall efficiency of such devices are the thickness of the active film and the modulation of the refractive index inside it.
VPHG technology has been applied in different astronomical spectrographs, at room and cryogenic temperatures \citep{wiyn, andrea7, andrea3, lepine, hou, andrea1, andrea9}. They have also been used as tuneable filters and cross dispersers \citep{mendes, gibson, castilho}. The VPHGs assembled in a GRISM configuration have been also used in this kind of instrumentation.
The availability of a low resolution grism covering the red part of the optical spectrum is of paramount importance for several astrophysical targets, among which stands out the study of supernovae in general and type Ia supernovae in particular. Such a grism will characterise the evolution of important lines during the photospheric (e.g. O I 7774 $\textrm{\AA}$ and Ca II IR triplet), and the nebular phases (e.g. Ca II] 7291-7323 $\textrm{\AA}$ and Ca II IR triplet).\\
Particularly important will be to study the evolution in type Ia supernovae of the high velocity features (HVFs) seen in the Ca II triplet profile, especially at early, pre-maximum, phases \citep[see][for a recent review]{chil13}. The origin of the HVFs remains unknown, but different hypotheses have been proposed, which include the impact of the ejecta with a circumstellar shell lost by the progenitor before the explosion \citep[see][]{ger2004}; an enhancement in the abundance of intermediate-mass elements (IMEs) in the outermost layers of SN Ia ejecta \citep{mazzali05a, mazzali05b, tanaka08}; or variations in the ionisation state of IMEs in the outer layers of SN Ia ejecta \citep{blondin2013}.\\
It is evident that the study of the Ca II HVFs could have a big impact in deriving the true progenitor scenario involved in the SNIa explosion, which in turn could have important implication in the use of SNIa in Cosmology. Moreover, the circumstellar (CS) material responsible for the HVFs could cause subtle alteration of the spectral energy distribution of SNIa, which could have possible consequences in the luminosity standardization of SNIa \citep{chil13}.
\\
In order to accomplish the desired requirements we designed and manufactured a VPHG based on a completely new holographic material. 
The study of this new material, other than dichromate gelatins (DCGs), is very important in order to make possible the design and manufacturing of innovative and large VPHGs.
Indeed DCGs are difficult to handle, they require a complex chemical process and the scalability to very large size gratings can be an issue. For this reasons we focused the attention to solid photopolymers  which combines high throughput, high refractive index modulation, self developing (i.e. no chemical processes needed) and size scalability. Such new material belongs to the class of solid photopolymers and this is the first time, in our knowledge, that this kind of holographic material has been used to make scientific grade dispersing elements. In the past only liquid photopolymers have been used once to produce VPHGs mounted in MOIRCS and FOCAS \citep{andrea4, andrea6}.
Starting from the scientific case of the SN 2013fj hereinafter we demonstrate the capabilities in terms of spectral resolution and throughput of this new family of Volume Phase Holographic Grating, based on  photopolymers, comparing two spectra of SN 2013fj in which one of them is secured by the adoption of previous existing state of the art grating.
Throughout this paper we refer to this new grating as VPH6.

\section{Observations and data reduction}
\begin{figure}[htbp]
  \centering
  \resizebox{\hsize}{!}{\includegraphics{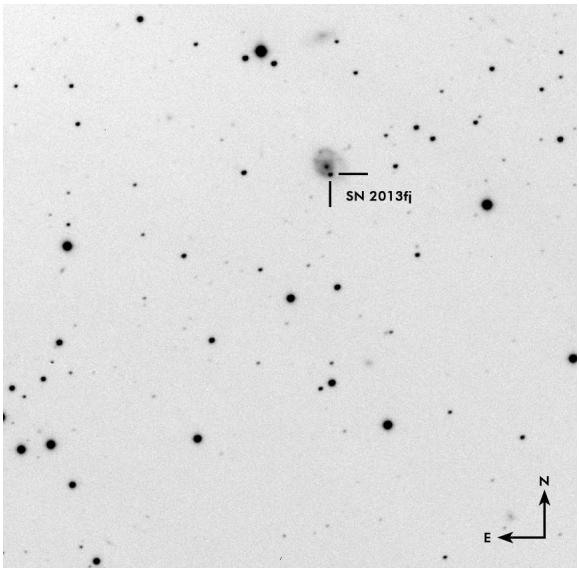}}
  \caption{R Band image of FoV around SN 2013fj. The plate scale of the image is 18.59$^{\prime\prime}$ px$^{-1}$. Exposure time is 120s. Seeing during the observation, taken at airmass 1.17 and measured on the image of the field, is 2.2$^{\prime\prime}$. The host galaxy of SN 2013fj is also clearly visible.}
  \label{fig:Snfield}
\end{figure}

We obtained the spectrum of SN 2013fj in visitor mode at Ekar Asiago Observatory using the Asiago Faint Object Spectrograph Camera (AFOSC) on 13th September 2013. The seeing during the night was quite constant (2.0 - 2.2$^{\prime\prime}$) and the sky was almost clear during the observations. In order to compare the new VPH6 device performance, we took two spectrum of the SN 2013fj. The first one was been obtained configuring the instrument with the GRISM GR04, yielding a dispersion of $\sim$ 5 $\textrm{\AA}$ px$^{-1}$ and R $\sim$ 600 in the spectral range 3500-7500 $\textrm{\AA}$. The spectra obtained with the new VPH6 GRISM, which was been secured immediately after the previous one, yields a dispersion of $\sim$ 3.5 $\textrm{\AA}$ px$^{-1}$ and R $\sim$ 500. We adopted a slit of $1.69^{\prime\prime}$ $\times$ 5.00$^{\prime\prime}$ for both spectra.\\
Since the Supernova Program has many observation priorities scheduled for AFOSC, we decided to first secure a spectrum with the well known GR04 grating (this was done to guarantee the data for the required scientific tasks). After that we coped to obtain another observation of the same target (SN 2013fj), under the same sky conditions, reducing the telescope time that would be otherwise not allocated for the other targets in the night.
%\textbf{Since the Supernova program is critical for AFOSC, we decided to first secure a spectrum with the well known GR04 grating, this aimed at guarantee a spectrum for scientific tasks required; then we coped to obtain another observation of the same target (SN 2013fj), under the same sky conditions, reducing the telescope time that would be otherwise not allocated for the other targets in the night.}
%Since the Supernovae program at AFOSC is time critical, we decided to secure a spectrum of the target first with the well known GR04 in order to guarantee basic scientific results strictly related to the supernova. Then a second spectrum has been obtained with the new VPHG adopting a slightly lower exposure time, aimed both at reducing overhead nightime and maintaing same observation conditions in terms of sky clearness and seeing. 
The integration time for the spectrum obtained with VPH6 was 1200s while for the GR04 was 1800s.\\
For each exposure we reduced data adopting standard IRAF\footnote{IRAF (Image Reduction and Analysis Facility) is distributed by the National Optical Astronomy Observatories, which are operated by the Association of Universities for Research in Astronomy, Inc., under cooperative agreement with the National Science Foundation.} procedure. We performed bias subtraction and flat field correction for each scientific frame adopting calibration obtained in the same night. The wavelength calibration was achieved using the spectra of standard arcs (Th-Ar and Hg-Cd) while flux calibration has been assessed through relative photometric calibration of standard stars spectra \citep{oke1990} obtained in the same night (BD+33d2642). The accuracy on wavelength calibration is $\sim$ 0.5 $\textrm{\AA}$ rms for the VPH6 and $\sim$ 0.2 $\textrm{\AA}$ rms for the GR04.\\ 
The two RMS values for the accuracy on the wavelength calibration are quite different since in the red part of the spectrum few comparison lines are available due to the calibration lamps installed in the AFOSC spectrograph. The calibration is more accurate in the blue because more emission lines in that part of the spectrum were usable. For this reason, the two RMS are slightly different. However, it is also important to note that the two values are far below the resolution power of the two gratings.
The apparent R magnitude obtained was 17.2 $\pm$ 0.2. We cross checked the calculated value through an aperture photometry of the R band acquisition image of the field (see Figure \ref{fig:Snfield}), secured just before obtaining the spectra.

\label{sec:observations}
\subsection{The device VPH6 at AFOSC}
The GRISM consists in a photopolymer based Volume Phase Holographic Grating (VPHG) (see Table \ref{tab:grism} for the GRISM features and requirements). The solid photopolymer used in this device has been recently developed by Bayer MaterialScience AG (product family: Bayfol$^{\textregistered}$ HX) as high performance holographic material \citep{bayer1}, for reflection holograms. The material is also suitable for transmission holograms with high dynamic range and sensitivity. Moreover the material is laminated onto flexible and transparent substrates of large sizes.
The grating is a "low dispersion" device with a line density of 285 lines mm$^{-1}$ with a target efficiency of 90\% at the central wavelength.

\begin{small}
\begin{deluxetable}{ c c c l l }
\tabletypesize{\scriptsize}
\tablecaption{AFOSC's GRISM VPH6: main specifications and requirements}
\tablenum{1}
\label{tab:grism}
\tablehead{\colhead{$\lambda_{central}$ [nm]} & \colhead{lines mm$^{-1}$} & \colhead{$\Delta \lambda$ [nm]} & \colhead{$\eta_{peak}$} & \colhead{$\eta_{side}$} \\
\colhead{(1)} & \colhead{(2)} & \colhead{(3)} & \colhead{(4)} & \colhead{(5)}} 
\startdata
  800 & 285 & 620 - 980 & 90\% & 30\% \\
\enddata
\tablecomments{Description of columns: (1) Working central wavelength of the grating; (2) Pitch of the grating; (3) Wavelength range; (4) 1-st order diffraction efficiency at the peak wavelength; (5) 1-st order diffraction efficiency at the wavelength range edges.}
\end{deluxetable}
\end{small}

\begin{figure}[htbp]
  \centering
  \resizebox{\hsize}{!}{\includegraphics{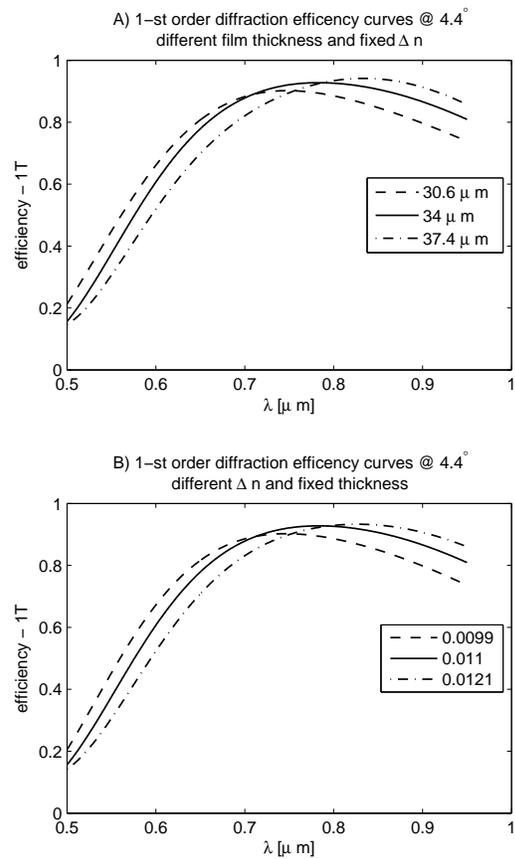}}
  \caption{Simulated 1-st order diffraction efficiency curves, with Rigorous Coupled Wave  Analysis; A) curves for different thickness (d = 34 $\mu$m $\pm$ 10 \%) and fixed $\Delta$n = 0.011; B) curves for different refractive index modulation ($\Delta$n = 0.011 $\pm$ 10 \%) and d = 34 $\mu$m . The considered incidence angle is 4.4$^{\textdegree}$ (on the grating interface).}
  \label{fig:simulation1}
\end{figure}

It can be seen that this low dispersion grating, combined with a suitable wavelength range, allows to covers the H$\alpha$ region and the Ca I bump typical of SN spectra.
The design of the grating was aimed at finding the best key parameters (film thickness and refractive index modulation) matching the scientific requirements, i.e. diffraction efficiency, wavelength coverage, and resolution. This activity has been performed through RCWA simulations\footnote{RCWA code, written in C, was provided by Gary Bernstein, who implemented the methods of Moharam \& Gaylord.} \citep{rcwa}.
We have therefore identified the best couple of parameters ($\Delta$n = 0.011 and d = 34 $\mu$m). The quite large thickness and small modulation of the refractive index is chosen in order to reduce the efficiency in orders higher than the first, which is a common feature of low line density VPHGs.
In Figure \ref{fig:simulation1} are reported the simulated 1-st order diffraction efficiency curves for different values of $\Delta$n and film thickness.
In the inset A) are reported the curves at different film thickness with with a fixed value of $\Delta$n = 0.011. Shown curves have $\pm$ 10 \% from the chosen value of 34 $\mu$m.
In the inset B) are reported the curves at different $\Delta$n with with a fixed thickness d = 34 $\mu$m. Shown curves have $\pm$ 10 \% from the chosen value of 0.011.
The photosensitive film was laminated onto a BK7 substrate before the exposure; the writing procedure was accomplished using a standard two-beams holographic setup with a DPSS laser of 532 nm.
The target refractive index modulation has been reached optimising the writing laser power since the final achieved $\Delta$n strongly depends upon the exposure power density as reported by \cite{bayer2}.

\begin{figure}[htbp]
  \centering
  \resizebox{\hsize}{!}{\includegraphics{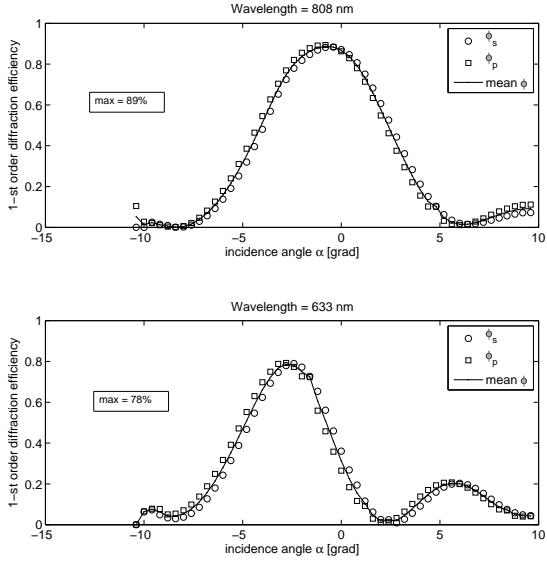}}
  \caption{Measured 1-st order diffraction efficiencies of the GRISM at 808 nm and 633 nm. The data represent the p-polarisation efficiency $\phi_p$, the s-polarisation efficiency $\phi_s$ and the resulting mean $\phi$, as function of the incidence angle $\alpha$ in air.}
  \label{fig:eff2}
\end{figure}

Regarding the GRISM, the apex prism angle has been designed in order to maintain the 1-st order central undieviated wavelength at 800 nm. The result was a BK7 prism with an angle of 12.7$^{\textdegree}$ that correspond at an entrance angle in the grating of 4.4$^{\textdegree}$.   
\\
The grating was then coupled with the prisms using a refractive index matching oil.  
In order to characterise the device, we measured the diffraction efficiency of the GRISM. The measurements were carried out using laser light at different wavelengths, setting the p- or s- polarisation and collecting the efficiency of the first order as function of the incidence angle;
two efficiency curves are reported in Figure \ref{fig:eff2}.

\begin{figure}[htbp]
  \centering
  \resizebox{\hsize}{!}{\includegraphics{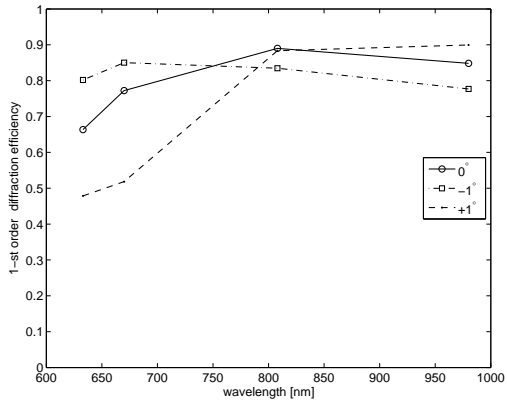}}
  \caption{Measured 1-st order diffraction efficiency curve of the aligned GRISM at different incidence angles. The angle 0$^{\textdegree}$ refers to the perpendicular to the grating inside the GRISM.}
  \label{fig:effblaze}
\end{figure}

It can be seen that the efficiency is very high even with the prisms coupled with the VPHG in the GRISM structure. This is achievable thanks to the  AR-coating onto the prisms surfaces and the use of the same substrate material for both the prisms and grating windows, that avoids further reflection losses.
We lastly report in Figure \ref{fig:effblaze} the measured 1-st order diffraction efficiency vs. wavelength of the GRISM aligned and mounted in his housing.

It is clear that the final alignment is crucial to maintain the requirements satisfied since even a tilt of a few degrees have a huge impact on the efficiency curve.

\section{Results}

\label{sec:result}

\begin{figure}[htbp]
  \centering
  \resizebox{\hsize}{!}{\includegraphics{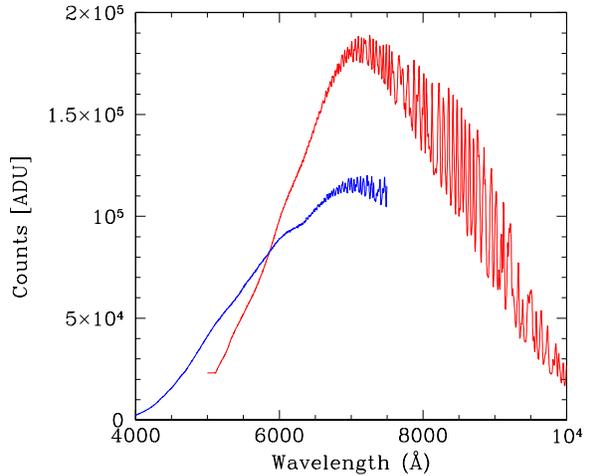}}
  \caption{Comparison of flat fields of VPH6 (red line) and GR04 (blue line) obtained at AFOSC adopting homogeneous configuration of the system.}
  \label{fig:flat}
\end{figure}

After the commissioning of the new VPH6 for AFOSC at 1.82m Copernico telescope, we taken two exposures of a flat field lamp with the same exposure time but varying the gratings in order to have a sound comparison of the two overall efficiencies. In particular, as shown in Figure \ref{fig:flat} the spectra of the lamp represents the behaviour of the compared gratings. 

Thanks to the homogeneous configuration of the instrument, in terms of efficiency of the detector, slits loss and telescope throughput, adopted for the two exposures it is possible to infer that starting from $\sim$ 5800 $\textrm{\AA}$ the number of counts in the spectrum of the new device (red curve) is significantly higher than those obtained in the case of the GR04 (blue curve).
Moreover the spectral coverage of the VPH6 is extended up to $\sim$ 9500 $\textrm{\AA}$ as required in order to investigate near IR properties of the targeted object for the SN program 
\citep{tomasella}.\\
The well visible fringing of the flat field recorded with the VPH6 was compared with the one of the GR04 after the signal normalisation; the result was a comparable fringe pattern, reassuring us that the effect can be attributed only to the CCD as described in the AFOSC's manual. \footnote{The entire manual of AFOSC is available at \url{http://archive.oapd.inaf.it/asiago/5000/5100/man01_2.ps.gz}}

\begin{figure*}[htbp]
  \centering
  \resizebox{\hsize}{!}{\includegraphics{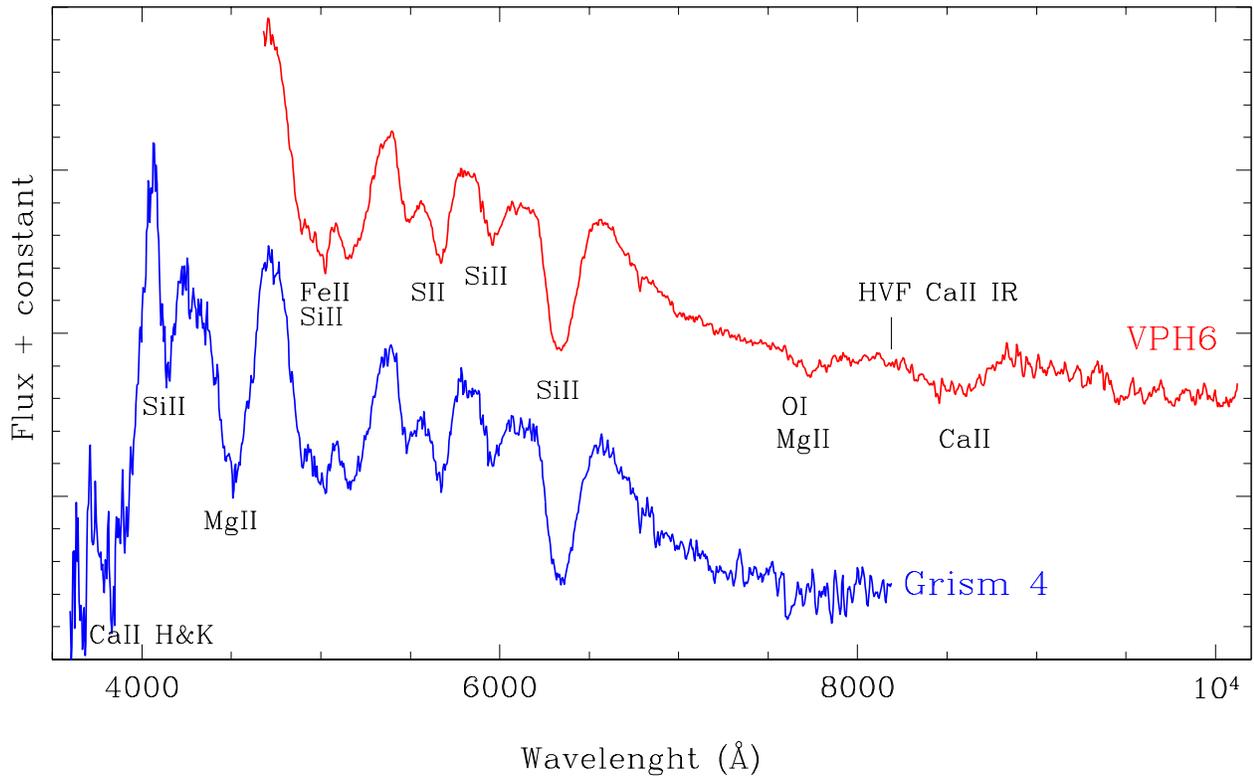}}
  \caption{Spectra of SN 2013fj taken with AFOSC and GR04 (blue line) and with VPH6 (red line). The principal lines of Si II, Ca II, S I, Mg II and Fe II are shown.}
  \label{fig:spec_13fj}
\end{figure*}

\begin{figure}[htbp]
  \centering
  \resizebox{\hsize}{!}{\includegraphics{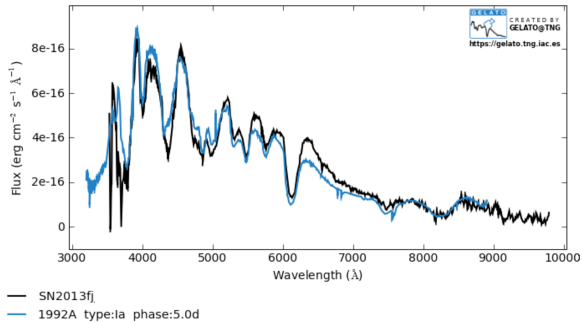}}
  \caption{Comparison of the SN 2013fj spectrum with that of phase +5 days from B maximum of SN 1992A made using the GELATO tool; GEneric classification Tool by \cite{Harutyunyan}. GELATO is a software for objective classification of Supernova spectra and performs an automatic comparison of a given spectrum with a set of well-studied SN spectra templates. Spectra of all SN types from Padova-Asiago SN Archive are used as templates for the comparison procedure.}
  \label{fig:conf}
\end{figure}

In order to assess the capabilities of the new grating in terms of scientific goals, we performed observations of SN 2013fj that has been already discovered by the amateur astronomers \cite{ciabattari} members of the Italian Supernovae Search Project on 7th September 2013. The spectra reported in Figure \ref{fig:spec_13fj} has been obtained 6.07 days later and is that typical of a Type Ia supernova, about 5$\pm$2 days after maximum light (see Figure \ref{fig:conf}), confirming the estimated phase reported by \cite{zanutta01}, and derived from a fast reduction performed at the telescope of the same data.\\
The recorded spectrum, reported in Figure \ref{fig:conf} and analysed with GELATO, is the combination of the spectra taken with the new VPH6 and the 3500-4750 {\AA} region of the GR04 (since they well overlap in the whole common range). This decision has been made because of the evident better Signal to Noise Ratio of the VPH6 and the higher coverage in the red region where the the important features of the target object are (such as HVF Ca-II IR).
The spectrum exhibit the broad P-Cygni lines typical of SNe Ia: the characteristic deep absorption near 6150 \AA\/ due to Si II 6347, 6371 \AA\/ (hereafter Si II 6355 \AA\/), the Si II 5958, 5979 \AA\/ feature (hereafter Si II 5972 \AA\/), the W-shaped feature near 5400 \AA\/ attributed to S II 5468 \AA\/ and S II 5640 \AA\/. \\ Other prominent features are Ca II H$\&$K, Mg II 4481 \AA\/, and several blends due to Fe II and Si II. At red wavelengths, particularly strong features are the Ca II near-IR triplet. Despite contamination from the 7600 \AA\/ telluric feature, O I 7774 \AA\/ is clearly visible.\\
Adopting for the host galaxy (CGCG 428-62) of SN 2013fj a recessional velocity of 10064 km s$^{-1}$, \cite{huchra}, an expansion velocity of about 10700 km s$^{-1}$ is deduced from the Si II 6355 \AA\/ absorption, while from the Ca-II IR triplet minimum an expansion velocity of about 11500 km s$^{-1}$ is deduced (the velocity is relative to the average Ca-II IR triplet wavelength, 8579.1 \AA\/). This behaviour is typically seen in SNIa, where the strong CaII lines are formed well above the photosphere, which is better traced by the weaker S-II lines (from which a mean expansion velocity of about 8400 km s$^{-1}$ is deduced).\\
The CaII-IR high velocity feature is by this phase very weak, see \cite{mazzali05b}, but possibly still visible as a weak absorption at about 23000 km s$^{-1}$\\
The expansion velocity deduced from the S-II 6355 \AA\/ minimum most probably places SN 2013fj among the low velocity gradient type Ia supernovae, following \cite{benetti1}.
\\
In order to make a sound comparison between the two different dispersive elements, we evaluated the S/N (Signal to Noise Ratio) at 6 different wavelengths along the two obtained spectra.
The results are reported in Table \ref{tab:SN} where S/N of the GRISMs have been normalised adopting the usual S/N equation \citep{SNeq} taking into account the different exposure times.\footnote{For the renormalisation we assumed that the equation simplifies to $S/N=\sqrt{s}$ (where $s$ is the signal from the source) since other noises (such as dark current or detector readout noise) are negligible at this level of comparison.}\\

\begin{small}
\begin{deluxetable}{ c c c }
\tabletypesize{\scriptsize}
\tablecaption{Signal to Noise Ratio comparison between the two GRISMs}
\tablenum{2}
\label{tab:SN}
\tablehead{\colhead{Wavelength [\AA]} & \colhead{S/N of VPH6} & \colhead{S/N of GR04} \\} 
\startdata
5000 &  22 &  17   \\
5500 &  42 &  27   \\                     
6000 &  45 &  27   \\
6500 &  57 &  27   \\
7000 &  38 &  20   \\
7500 &  30 &  13   \\
\enddata
\tablecomments{The S/N of the two GRISMs have been normalised according to the respective exposure times.}
\end{deluxetable}
\end{small}

\section{Conclusions}
\label{sec:discussion}

We demonstrated the good performances obtainable by using a Volume Phase Holographic Grating based on new photopolymer materials which are self developing, characterised by a high sensitivity and dynamic range in conjunction with an easy processability. The GRISM has been designed and manufactured in order to maximise the efficiency reducing the reflection losses. 
For these reasons such devices are a reliable alternative to classical VPHGs.
We assessed the scientific requirements which drawn the design of this new DOE by collecting the spectrum of the newly discovered SN Ia PSN J22152851+1534041 = SN 2013fj . We finally carried out a sound comparison with another spectrum of the same object under the same conditions, secured with the standard grism that is characterised by the same dispersion and resolution.
\\
\\
\textbf{Acknowledgements}
\\
We are grateful to:
Dr. Thomas F\"{a}cke of Bayer MaterialScience for providing the material and for the useful discussions; 
The technicians at mt. Ekar for all the support during commissioning based on observations collected at Copernico telescope (Asiago, Italy) of the INAF - Osservatorio Astronomico di Padova;
L.T. and S.B. are partially supported by the PRIN-INAF 2011 with the project "Transient Universe: from ESO Large to PESSTO".

\end{document}